# Resolving the intrinsic short-range ordering of K$^+$ ions on cleaved muscovite mica

G. Franceschi*,1, P. Kocán2, A. Conti1, S. Brandstetter1, J. Balajka1, I. Sokolović1, M. Valtiner1, F. Mittendorfer1, M. Schmid1, M. Setvín1,2, and U. Diebold1

1 Institute of Applied Physics, TU Wien, Wiedner Hauptstraβe 8-10/E134, 1040 Vienna, Austria

2 Department of Surface and Plasma Science, Charles University, V Holesovickach 2, 180 00 Prague, Czech Republic

Corresponding Author *[franceschi@iap.tuwien.ac.at](franceschi@iap.tuwien.ac.at)

**Abstract**

Muscovite mica, $KAl_2(Si_3Al)O_{10}(OH)_2$, is a common layered phyllosilicate with perfect cleavage planes. The atomically flat surfaces obtained through cleaving lend themselves to scanning probe techniques with atomic resolution and are ideal to model minerals and clays. Despite the importance of the cleaved mica surfaces, several questions remain unresolved. It is established that $K^+$ ions decorate the cleaved surface, but their intrinsic ordering – unaffected by the interaction with the environment – is not known. This work presents clear images of the $K^+$ distribution of cleaved mica obtained with low-temperature non-contact atomic force microscopy (AFM) under ultra-high vacuum (UHV) conditions. The data unveil the presence of short-range ordering, contrasting previous assumptions of random or fully ordered distributions. Density functional theory (DFT) calculations and Monte Carlo simulations show that the substitutional subsurface $Al^{3+}$ ions have an important role for the surface $K^+$ ion arrangement.

**Introduction**

The importance and popularity of muscovite mica in surface and interface science can hardly be exaggerated. Over the last decades, the surface of this material has been the focus of hundreds of theoretical and experimental studies across diverse fields[1,2] including environmental science, bio- and geo-chemistry, nanotribology, new-generation electronics based on 2D materials, and thin-film growth. Muscovite mica (mica, hereafter) is a common phyllosilicate with a nominal composition of $KAl_2(Si_3Al)O_{10}(OH)_2$ and a layered structure with alternating aluminosilicate and $K^+$ layers (Figs. 1a, b). Each aluminosilicate layer is made of three sheets: one with $AlO_6$ octahedra and OH groups (octahedral sheet), sandwiched by two sheets of 75% $SiO_4$ and 25% $AlO_4$ tetrahedra (tetrahedral sheet). The tetrahedra in the tetrahedral sheet form a distorted hexagonal structure, where each ditrigonal cavity (ring) hosts one $K^+$ ion that compensates for the formal −1 charge introduced by the substitutional Al ions. The material easily splits apart at the K$^+$ layers, which leaves half of the K$^+$ ions on each created surface and maintains charge neutrality.

To the surface scientist's delight, cleaved mica surfaces are atomically flat and virtually free of steps[3]. This, together with the broad available knowledge of its bulk properties[1], has made mica a widely used model substrate in many contexts. Prominent examples are studies on the adsorption and dynamics of biomolecules[4–7], motivated by the suggestion that life originated between mica sheets[4], and studies on water[8–13], to model the atomic-scale mechanisms underlying the water-mineral interaction that is ubiquitous on the Earth. On a more technical side, mica's facile preparation and flatness have made it the test system of choice for emerging experimental techniques with real-space, molecular resolution in ambient or liquid conditions beyond the traditional atomic force microscope (AFM), such as the 3D AFM[14–17].

Despite the vast popularity of mica, open questions about the system remain. Its surface K$^+$ ions are central to many studies, as they can be easily exchanged in solution[10,18–23] and offer an exciting playground to investigate ion hydration[8,24–27] and ice nucleation[12,13,28] on mineral surfaces. Yet, the intrinsic K$^+$ distribution at the surface – unaffected by the interaction with the environment – is unknown due to the absence of UHV direct imaging. AFM images of the mica surface have been obtained in air or solution[15,24–26,29–34], showing the cation hydration structures in these environments. However, these arrangements do not necessarily correspond to the intrinsic ones. In the ambient, the ready adsorption of water and airborne impurities[2,35] modifies the ion-ion interaction and may promote their mobility[33], in turn modifying the ions' arrangements. In solution, the distribution of the hydrated ions may be affected by their increased mobility[20], ion-water and water-water interactions[24], and the pH. The measured arrangements have been explained in terms of water-mediated ion-ion interactions[24,33], while the potential role of the aluminosilicate subsurface has not been considered or deemed negligible[24]. As shown in this work, however, this assumption should be revised.

Doubts exist not only on the surface K order but also on the Al order in the subsurface tetrahedral sheets. The Al distribution is hard to determine experimentally since Al and Si have similar scattering factors in X-ray diffraction. Early nuclear magnetic resonance (NMR) combined with Monte Carlo (MC) simulations[36,37] have suggested the presence of Al short-range ordering. Such ordering could affect the distribution of the surface $K^+$ ions through electrostatic interaction. Testing this hypothesis in the ambient or solution is, however, difficult for the reasons listed above.

Imaging the mica surface under ultra-high vacuum (UHV) should be well suited to assess the intrinsic ordering of the $K^+$ ions and potentially relate it to the distribution of the subsurface Al ions. However, so far, individual $K^+$ ions could not be resolved because UHV cleaving often introduces strong electrostatic fields that make AFM imaging challenging[38]. To the best of the authors' knowledge, the only account of $K^+$ ordering after UHV cleaving comes from low-energy diffraction (LEED)[39], which has suggested a random distribution. Instead, the present results – based on non-contact (nc) AFM acquired on UHV-cleaved, clean mica – show that its surface $K^+$ ions are arranged with short-range order. The distribution is analyzed with density functional theory (DFT) calculations and MC simulations, which demonstrate a close relation to the subsurface Al arrangements.

## Results and Discussion

Figures 1c, d show atomically resolved images of the mica surface after UHV cleaving. The images were acquired using the qPlus sensor[40,41], which is stiffer (2000−3500 N/m) than standard

AFM cantilevers. Hence, it is less affected by the long-range interactions with the highly charged surface of mica[42,43] that otherwise hamper atomic contrast[38]. The images reveal an array of isolated, round, dark features arranged on a hexagonal lattice. The dark contrast represents attractive interaction between the nc-AFM tip and the sample (negative frequency shift). In the background, regions of different contrast of a few nanometers in width are visible (two of them are marked by a white and a black circle in Fig. 1c). The background contrast variation suggests a different long-range interaction with the AFM tip, possibly originating from trapped subsurface charges. This is also evidenced by frequency shift curves acquired as a function of tip-sample distance and sample bias (Supplementary Note 4).

The isolated dark features in Fig. 1 are assigned to the $K^+$ ions left on the surface after cleaving. The features sit on a hexagonal lattice with the expected lattice constant of 0.52 nm, as determined by the Fourier Transform of Fig. 1e (see Supplementary Note 5 for the analysis of the diffuse background of the Fourier transform). As expected from electrostatic considerations, the cations occupy approximately half (precisely 47.8±0.1%) of the surface sites. Supplementary Note 5 discusses how the coverage was derived and why it differs from the expected 50%. It is worth noting that the exclusively attractive interaction between isolated, undercoordinated cationic adatoms on the surface and the tip is typical in nc-AFM[44,45]. The $K^+$ ions cannot be deliberately nor inadvertently manipulated with the AFM tip, and their relatively large height hampers the resolution of the underlying aluminosilicate sheet.

An x-ray photoelectron spectroscopy (XPS) analysis further supports the assignment of the dark features to $K^+$ ions. The XPS survey and the C 1$s$ region in Supplementary Fig. 5 show that UHV-cleaved mica is clean except for typical substitutional trace impurities (Fe, Mg, and Na)[20,35] listed in the supplier's datasheet. Moreover, comparing the K 2$p$ spectra acquired in normal and grazing emission (Figs. 2a, b, respectively) points to the presence of K at the surface. The marked difference in the shape of the profiles (less pronounced separation in grazing emission) is rationalized considering that XPS simultaneously probes two types of $K^+$ ions: the three-fold-coordinated surface ions and the six-fold-coordinated bulk ions. Because of their different coordination, these species appear as two spin-orbit doublets with different core-level shifts. The surface component is more pronounced in grazing emission (≈43.5% of the total peak area) than in normal emission (≈11%) because of the larger surface sensitivity of the former acquisition mode. The measured separation between bulk and surface components, 1.22 eV, is reproduced by the calculated initial-state difference of 1.14 eV between the core levels originating from the surface and bulk $K^+$ ions.

A small fraction of the dark species in Figs. 1c, d appear with slightly different size and contrast than the rest (arrows highlight examples). Such contrast variations are expected if atoms are located at different heights or if they are different chemical species. As discussed in Supplementary Notes 1 and 5, the species with increased contrast could correspond to $K^+$ ions sitting on defective aluminosilicate rings characterized by 3 Al ions (protruding ≈0.1 Å more than $K^+$ ions sitting on regular sites, according to DFT); they could also be $Ca^{2+}$ substitutional impurities, whose higher charge compared to the $K^+$ ions should cause a stronger interaction with the AFM tip. On the other hand, the fainter species are likely to be $Na^+$ ions, characterized by a smaller ionic radius than $K^+$ ions.

The high resolution of the nc-AFM images in Figs. 1 and 3a affords detailed insights into the surface $K^+$ ion ordering. The ions arrange with short-range order, forming alternating rows along the three low-index directions with an average length of 3.5±0.4 nearest neighbors (NN), exemplified by the black lines in Fig. 3a. These rows are often interrupted or joined by 120° kinks of three NN, some of which are marked in yellow in Fig. 3a. The short-range order is quantified by the autocorrelation analysis in Supplementary Note 5.

What is the origin of the measured short-range order of the surface $K^+$ ions? It is natural to expect that the repulsion between the $K^+$ ions plays an important role. Another expected contribution is the electrostatic interaction between surface $K^+$ and subsurface $Al^{3+}$ ions: $K^+$ should prefer $Al^{3+}$ neighbors vs. $Si^{4+}$ neighbors because of reduced electrostatic repulsion. This picture is confirmed by the DFT calculations and MD simulations discussed below.

The DFT calculations were performed with the Vienna Ab-initio Simulations Package (VASP)[46,47] using the r$^2$SCAN metaGGA[48] exchange-correlation functional (see Methods for computational details). This functional describes well the structural properties of bulk mica (lattice constants and angles deviate less than ~0.6% from the experimental values) and yields an improved value of the bulk modulus (54 GPa) compared to the GGA (PBE) functional used in previous studies[49,50]. Several models were tested, each characterized by different arrangements of $K^+$ ions at the surface or $Al^{3+}$ ions in the upper $AlSi_3$ tetrahedral sheet (see Figs. 3b−e for a selection, and Supplementary Fig. 1 for the complete set). The $K^+$ ions were placed in straight or zigzag rows, reproducing the preferred distributions observed experimentally (Fig. 3a). Each structure is further characterized by the arrangement of the substitutional $Al^{3+}$ ions in the subsurface rings (meta or para), and, additionally, by their number and positions in the rings occupied by a $K^+$ ion (see insets). The relative energies per $K^+$ ion referred to the lowest-energy structure are shown at the bottom of each panel. Two main observations emerge from DFT: (i) The most stable $K^+$ rows and $K^+$ zigzag arrangements (Figs. 3b and 3d, respectively) are comparable in energy. Their small energy difference ($\Delta E \approx 60$ meV) is consistent with the coexistence of 120° kinks and straight rows observed in the experiments. (ii) Given a particular $K^+$ order, structures with 2 Al per K-occupied ring are always favored compared to those with only 1 Al. This is exemplified by the comparison between Figs. 3b and 3c ($\Delta E \approx 170$ meV) and between Figs. 3d and 3e ($\Delta E \approx 260$ meV). The latter observation hints at a strong link between $K^+$ and $Al^{3+}$ order, further supported by the MC simulations below.

DFT was further employed to calculate hopping barriers for the $K^+$ ions. The lowest calculated hopping barriers corresponded to $K^+$ ions jumping from 1-Al- to 2-Al-rings, i.e., towards lower-energy configurations. This justifies the assumption of the MC simulations that only consider diffusion towards lower-energy states (see below). The calculated diffusion barriers lie between 0.7 and 1 eV, corresponding to a time scale for hopping between less than a minute and a few hours at room temperature. Hence, at least some diffusion events should be allowed during the ≈3 minutes passing between room-temperature cleaving and sample transfer into the AFM cryostat held at 4.7 K. This entails that the observed $K^+$ distribution may not be uniquely determined by the interaction of the $K^+$ ions with the tetrahedral sheets above and below them before cleaving; if diffusion is allowed, the system can assume the minimum-energy configuration dictated only by the lower sheet after cleaving.

The link between the $Al^{3+}$ and $K^+$ order demonstrated by the DFT data was further explored by performing MC simulations (see Methods for the simulation details). The results are summarized in Fig. 4. The top row of Fig. 4 reports the MC-derived distributions of $K^+$ ions. The bottom row shows the corresponding histograms of the NN distributions (grey), always overlayed on the distribution obtained from the experimental data (dashed). Figure 4a shows the distribution obtained for an unsupported $K^+$ layer where the $K^+$ positions were distributed on a lattice with the same lattice constant as mica. In this simulation, only the $K^+-K^+$ electrostatic repulsion can affect the $K^+$ order. To obtain this distribution, the ions were initially randomly placed on the hexagonal lattice with 48% occupancy (the experimental concentration). Figure 4f shows the histogram of the initial (random) distribution, with a maximum at 3NN (black). Relaxing the system by MC produces a pronounced maximum at 2 NN due to the frequent alternating rows seen in Fig. 4a (within each row, each ion has 2 NN; this distribution is favored because it minimizes the number of NN ions, and, hence, the electrostatic repulsion). The experimental distribution also has a maximum at 2 NN. However, this maximum is less pronounced, indicating a somewhat lower degree of order. The difference in the experimental $K^+$ distribution compared to the simulated free-standing $K^+$ layer is consistent with the $K^+$ order being not uniquely determined by the $K^+-K^+$ interaction but also by the interaction with the underlying $Al^{3+}$ lattice, as indicated by DFT.

It is here useful to open a parenthesis on the knowledge available from the literature on the $Al^{3+}$ arrangement in the tetrahedral $AlSi_3$ layers of mica. Direct measurements are currently missing. X-ray diffraction data indicate the lack of long-range order but cannot and do not yield information about the actual arrangements[51]. $^{29}Si$ magic angle spinning (MAS)−NMR spectroscopy yields the probabilities of finding Si atoms in different tetrahedral environments (e.g., surrounded by 0, 1, or 2 Al atoms)[52]. Like the diffraction data, they do not allow on their own to determine the exact Al position in the rings (e.g., meta or para, when having 2 Al/ring) or the possible presence of short-range ordering. Previous MC simulations have determined possible distributions fitting the NMR data while obeying electrostatic constraints on the Al positions, specifically: Avoiding NN (Löwenstein's rule)[53] and having either 1 or 2 Al ions per ring to achieve maximum charge dispersion[36,52]. Among the tested distributions satisfying these constraints, the one best fitting the NMR data (in the following, referred to as the NMR fit distribution) had a 2:1 ratio of rings with meta and para configuration, consistent with meta and para being equally favored.

In the present work, MC simulations were performed to gain more information on the $Al^{3+}$ order based on the measured surface $K^+$ order (selected models in Figs. 4b−d; full set of simulated structures in Supplementary Fig. 2). The simulations were obtained starting from different subsurface $Al^{3+}$ models, including some incompatible with the NMR data[52] – such as the long-range-ordered and random distributions in Figs. 4b, c and Supplementary Fig. 2d – and the NMR fit distribution mentioned above (Fig. 4d). The simulations evidence one common trait, consistent with the DFT data: The K ions always follow the order of the underlying Al lattice, preferably occupying rings with 2 Al ions. As expected, the NMR-incompatible Al distributions (e.g., Figs. 4b, c) yield $K^+$ distributions that do not reflect the experimental data. More interesting, also the NMR fit yields an unsatisfactory agreement: In Figure 4i, the simulated $K^+$ distribution has too few $K^+$ ions with one and too many with three neighbors compared to the experiments. However, one must notice that the models considered up to now do not consider any screening effects through the substrate. In reality, screening effects are to be expected: When considering only NN interactions, analysis of the DFT results indicates that the K−K and the K−Al interactions are comparable in energy (≈0.15 eV and 0.17 eV, respectively), despite the larger distance of K−K neighbors (0.52 nm) compared with K−Al (0.36 nm). Hence, the screened K−Al interaction should be described well by a weaker charge of Al. When taking a lower effective charge of Al as an approximation for screening ($-0.65e$, instead of $-1e$), one can obtain a good fit with the experimentally observed K arrangements (Figs. 4e, j). The extreme situation, when almost all of the Al formal charge is screened by the substrate (imposed charge on Al = $-0.24e$) yields a similar distribution as the „free-standing" $K^+$ layer in Fig. 4a (Supplementary Figs. 2k, n). It is important to point out that the solution of Fig. 4e is not unique. Different ordering of the $Al^{3+}$ sublattice than assumed here could yield the same ordering of the $K^+$ ions and still be consistent with the NMR data (see Supplementary Note 2). Summarizing, the MC simulations confirm the important role of the subsurface $Al^{3+}$ ions for the $K^+$ order. They allow excluding many types of subsurface $Al^{3+}$ arrangements, but even when combined with the experimental data on the $K^+$ order, they do not yield a unique solution.

The short-range order reported here for the surface $K^+$ ions of UHV-cleaved mica furthers the current knowledge about the mica surface. To the authors' knowledge, the intrinsic surface $K^+$ order has not been determined directly in the previous literature. Early LEED works did not provide any evidence of ordering after UHV cleaving[39], in contrast with the short-range order reported here. In the present work, LEED could not be performed on the freshly cleaved samples because of charging. Patterns compatible with those in ref. [39] could be acquired on contaminated samples (see Supplementary Note 6). LEED is also unlikely to observe much short-range ordering on a lateral scale smaller than 10 nm[45]. Contrasting to the scarcity of UHV studies, more evidence about ion ordering on mica is found for surfaces in solution. AFM studies have shown that hydrated $K^+$ and $Rb^+$ ions[25,54] stabilized at mica-solution interfaces exhibit a similar ordering to that of the $K^+$ ions on the freshly cleaved surface reported here, i.e., preferably arranged in row segments rather than randomly. With support from molecular dynamics simulations, it was concluded[24] that the ordering was related to water-correlation effects, i.e., the energy gained by the global hydration structure when having a specific geometric arrangement of the ions, rather than to repulsion between the solvated ions or electrostatic interaction with the substrate. However, the similarity of the arrangements found in solution to those of the UHV-cleaved surface suggests that the interaction between ions and substrate may play a more significant role than previously assumed in determining the arrangement of (hydrated) ions on mica.

In conclusion, this study provides atomic resolution on the surface details of mica, a popular layered mineral in surface and interfacial science. Atomically resolved nc-AFM images after UHV cleaving reveal the intrinsic short-range ordering of its surface $K^+$ ions: These preferentially arrange in short, alternating rows, in contrast to previous assumptions of random arrangements. DFT calculations and MC simulations show that the $K^+$ ordering is not only due to the electrostatic repulsion between the $K^+$ ions but also strongly affected by the interaction with $Al^{3+}$ ions in the subsurface aluminosilicate layer. The atomic-scale insights provided by this study on UHV-cleaved mica broaden the current knowledge of mineral surfaces. They also offer valuable input to disentangle the many factors at play in the more complex ambient or liquid environments, where mica serves as a model system to unravel important atomic-scale processes.

## Methods
### Experimental methods

The experiments were carried out in a UHV setup consisting of two interconnected chambers: a preparation chamber for sample cleaving and XPS measurements (base pressure $< 1 \times 10^{-10}$ mbar) and an AFM chamber for nc-AFM measurements (base pressure $< 2 \times 10^{-11}$ mbar).

Natural muscovite mica single crystals [(0001) oriented disks of grade V1, with 10 mm diameter and 0.25 mm thickness, from TedPella – see Supplementary Fig. 5 for typical impurities] were glued on Omicron-style stainless steel sample plates with UHV-compatible epoxy glue (Epotek). They were cleaved in UHV at room temperature before each experiment. Two cleaving methods were used. The first consists of using a wobble stick to apply a tangential force to a metal stud glued on top of the sample[55]. The portion of the sample initially covered by the stud is thus cleaved, and it can be probed with nc-AFM. The second method uses a carbon-steel blade mounted on an Omicron-style plate to peel off a thin mica layer. This procedure induces the cleavage of an entire mica disk and can be repeated on a single sample several times. The latter technique was used for the presented XPS experiments. AFM data on the as-cleaved surfaces were acquired with both methods and yield identical surface structures.

XPS was performed with a non-monochromatic dual-anode Mg/Al X-ray source (SPECS XR 50) and a hemispherical analyzer (SPECS Phoibos 100). Spectra were acquired in normal and grazing emission (70° from the surface normal). Due to the insulating nature of the samples (bandgap of 7.85 eV)[56], the XPS spectra showed shifts to higher binding energies (between 5 and 7 eV). As already noted in the literature[35], the magnitude of the shift depends on the amount and type of surface contamination, XPS acquisition geometry, and sample thickness. For the display and analysis of the XPS data, an energy correction was applied to all spectra to set the peak value of the core-level K $2p_{3/2}$ to 293.75 eV reported in the literature[35]. This resulted in an O $1s$ peak at binding energy of 532.3 eV. The intensities and positions of the Al-Kα-excited XPS peaks were evaluated with CasaXPS after subtracting a Shirley-type background. The K $2p$ peaks were fit by two doublets, assuming an asymmetric Lorentzian line shape LA(1, 643). In each doublet, the relative peak separation was set to 2.8 eV in line with previous works[35], and the area ratio to 2:1. All peaks have the same FWHM. The separation between the two doublets was set the same for normal and grazing emission.

The AFM measurements were performed at 4.7 K using a commercial Omicron qPlus LT head and a differential cryogenic amplifier[57]. Frequency-modulated non-contact AFM mode was used. The tuning-fork-based AFM sensors ($k$ = 2000−3500 N/m, $f_0 \approx$ 45 kHz, Q $\approx$ 50000)[40] had a separate contact for tunneling current attached to electrochemically etched W tips that were cleaned *in situ* by field emission[58]. Before each measurement, the tips were further prepared on a clean Cu(110) single crystal by repeated indentation and voltage pulses. CO-functionalized tips[59] were used to image some samples. The coarse approach was done with a setpoint of −1 Hz. The controller was switched off, and the tip was gradually approached in the constant-height mode until an AFM contrast was visible while scanning in x and y. All AFM images presented here were acquired in constant-height mode. At times, the absolute values of frequency shifts obtained during the acquisition of atomically resolved images were large (up to 100 Hz) and were not reproducible on different regions on the same sample or on different samples. This is because cleaving can create domains of trapped charges that can cause long-range electrostatic interactions between the surface and the tip[38]. Short-range forces used for imaging the $K^+$ ions were in the order of 0.1 nN (attractive regime). The electrostatic fields created after cleaving can be partially compensated by applying a bias voltage between tip and sample. Most of the measurements were performed by applying a bias voltage such that the surface was measured as close as possible to the lowest local contact potential difference (LCPD), i.e., at tip-sample potential differences that were as close to the LCPD as possible. The bias voltage reported in the presented images corresponds to bias applied to the back of the sample plate while having the tip on ground. Supplementary Note 5 reports details about the statistical analysis of AFM images.

**Computational Methods (DFT)**

The DFT calculations were performed with the Vienna Ab-Initio Simulation Package (VASP)[46,47] using projector augmented wave (PAW) potentials[60] and the r$^2$SCAN metaGGA[48] exchange-correlation functional. The bulk structure was optimized with a cutoff energy of 1000 eV, and a k-point mesh of 4×2×1 was used to integrate the Brillouin zone. The optimized r$^2$SCAN lattice parameters ($a$ = 5.16 Å, $b$ = 9.00 Å, $c$ = 20.19 Å, and β=95.75°) are in excellent agreement with the experimental values ($a$ = 5.16 Å, $b$ = 8.95 Å, $c$ = 20.07 Å, and $\beta$=95.99°)[61] (maximal deviation of ~0.6%).

The surface calculations were performed with a lower cutoff energy of 500 eV and a 2×2×1 k-point mesh for the 2×1 unit cell (see Fig. 3). A single muscovite trilayer was used for the calculations, as double trilayer slabs yielded differences <10 meV/$K^+$ ion. The rows were formed along the [100] direction. Their rotation by 60° (tested because the surface does not possess an exact hexagonal symmetry) gave similar results, with differences <15 meV/$K^+$ ion. A symmetric setup was used for the distribution of the Al ions on both sides of the slab.

The diffusion barriers of the $K^+$ ions were determined by identifying the saddle points in the potential energy landscape with the improved dimer method[62], with the remaining forces at the saddle point smaller than 0.02 eV/A. This was followed by the relaxation of the $K^+$ ion from the saddle point to either the initial or the final state of the reaction step. Core-level shifts were determined in the initial state approximation.

**Simulations methods (Monte Carlo simulations)**

Electrostatic models consisting of a K layer sitting above an $AlSi_3$ layer were built inspired by the Metropolis algorithm[63]. Al ions were constrained to occupy ¼ of the lattice positions in the hexagonal rings of the $AlSi_3$ tetrahedral sheet. K ions were constrained to sit on a hexagonal lattice centered in the hexagonal rings of the $AlSi_3$ tetrahedral sheet, at a vertical distance of 1.85 Å from the underlying layer, according to the values derived from the DFT-optimized surface (see Supplementary Table 1). Unless noted otherwise, screening effects via the substrate were not accounted for.

To build the model, the subsurface $AlSi_3$ layer was generated within a circular area with a radius of ≈100 nm (Al ions treated as −1 point charges, Si ions as neutral). The full set of simulations in Supplementary Fig. 2 was obtained with different arrangements for the Al sites, as discussed in Supplementary Note 2. As a second step, the surface layer of K ions (+1 point charges) was added by randomly placing the charges on the same 100 nm-radius area of the AlSi$_3$ layer. A coverage of 48% was chosen to match the experimental findings. During the simulations, ions in a central area of ≈30 × 30 nm$^2$ above the subsurface area were allowed to move. The rest was frozen to avoid electrostatic expansion. After pre-calculating the field of the immobile ions, the simulation was run according to the following algorithm: Make a random jump of a randomly selected mobile K ion to an unoccupied neighboring site and accept the new configuration if its electrostatic energy has decreased. This procedure is motivated by the DFT-calculated barriers, which only allow energetically downhill diffusion at room temperature (see main text). The steps were repeated until reaching a local energy minimum, in which no successful jump appeared during 100 attempts per atom. The minimum was typically reached after 1.6−1.9 successful jumps per ion.

To compare the ion distributions in different models with the experimental distributions, histograms of the fraction of K$^+$ ions with a given number of nearest neighbors (NN) were taken as a metric (see, e.g., Figs. 4f–j). They represent the number of ions found with a given number of NN. Histograms of the experimental distributions were obtained from a point map of the ion positions, extracted as detailed in Supplementary Note 5.

## Data availability

The data that support the findings of this study are available from the corresponding author upon request.

## Acknowledgements

This work was supported by the European Research Council (ERC) under the European Union's Horizon 2020 research and innovation programme (grant agreement No. 883395, Advanced Research Grant 'WatFun'). M.V. and S.B. acknowledge support from the FFG Project FunPakTrio. I.S. and M.Se. were supported through the FWF project "Super". M.Se. and P.K. acknowledge the support from the Czech Science Foundation, project GACR 20-21727X. The computational results have been achieved in part using the Vienna Scientific Cluster (VSC).

## Author contributions

M. Setvin, M. V., and U. D. conceived the idea and started the project. G. F. acquired and analyzed the experimental data and wrote the manuscript. P. K. performed the MC simulations. A. C. and F. M. performed the DFT calculations. S. B. acquired nc-AFM images. I. S., J. B., and M. Setvin supported the acquisition of the experimental data. M. Schmid supervised the data analysis and interpretation. G. F. and U. D. supervised the project. All authors contributed to the discussions and the manuscript draft.

## Competing interests

The authors declare no competing interests.

# Figures

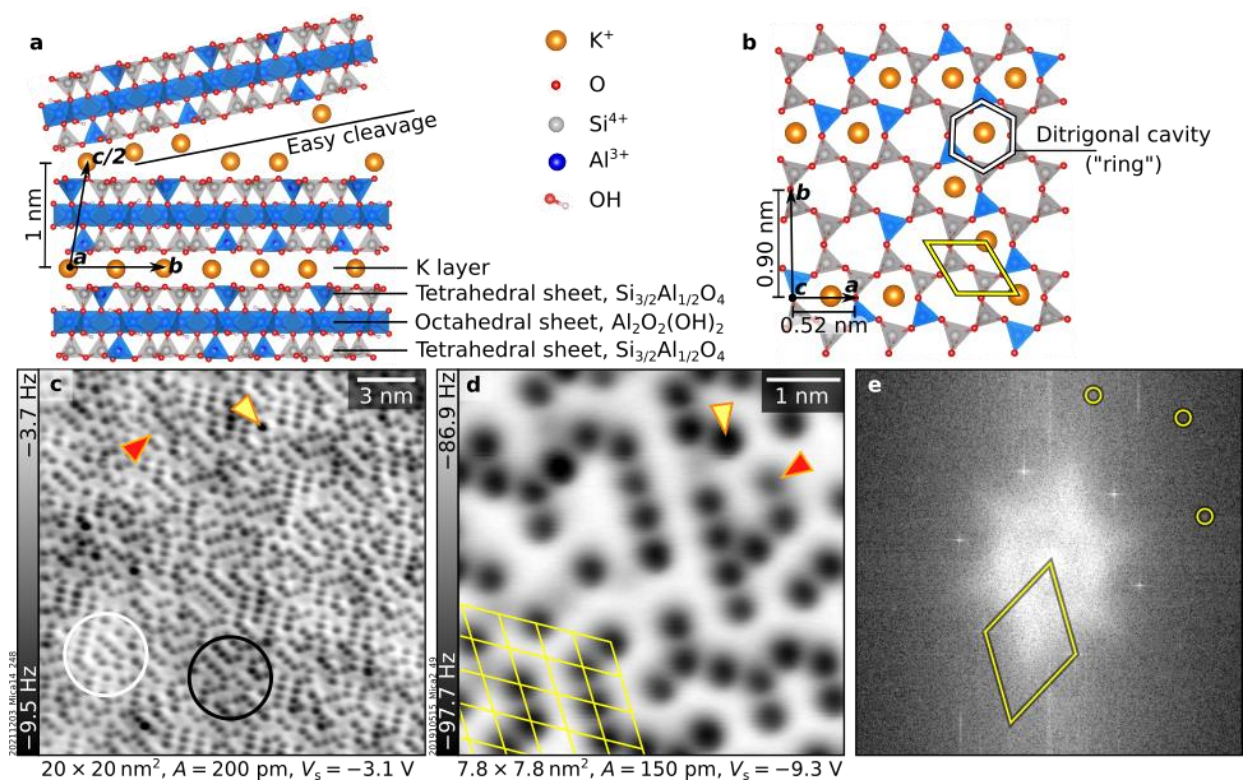

Figure 1. **Cation ordering on as-cleaved mica. a, b** Crystal structure of mica. Al ions (blue) in the tetrahedral sheets are placed in a pseudo-random arrangement akin to Fig. 4e, showing one possible arrangement fitting the experimental data. **a** Side view of bulk mica. Cleaving occurs at the K layer, leaving half the $K^+$ cations on each side. **b** Top view of the surface after cleaving. Before cleaving, each ditrigonal cavity (ring, highlighted in white) is occupied by one $K^+$ ion. After cleaving, 50% $K^+$ ions remain on each cleaved surface. **c, d** Atomically resolved constant-height nc-AFM images of mica after UHV cleaving, acquired with a CO-functionalized tip and a metal tip, respectively. The images were acquired at 4.7 K with different qPlus sensors and on different samples. Yellow (red) arrows highlight species with darker (fainter) contrast than average. **e** Fourier Transform of the image shown in panel **c**. Yellow circles mark selected diffraction spots of the underlying bulk. Unit cells of the (almost) hexagonal lattice in panels **b**, **d**, and **e** are highlighted in yellow (strictly speaking, the muscovite unit cell is rectangular because the tetrahedral rings are not perfect hexagons).

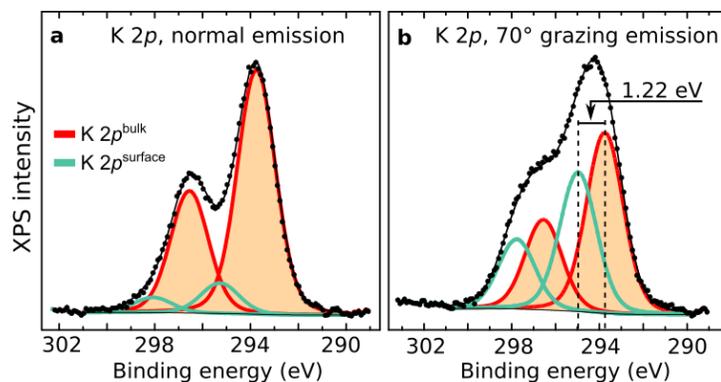

Figure 2. **XPS spectra of K$^+$ ions on UHV-cleaved mica**. XPS data (black points) and fitted curves (solid lines) of the K 2$p$ core levels (Al K$\alpha$, pass energy 20 eV). **a**, **b** Spectra taken in normal and 70° grazing emission, respectively. Two sets of K 2$p$ components fit the spectra, assigned to K residing in the bulk (orange) and on the surface (green). The binding energy axes were adjusted to account for charging (see Methods).

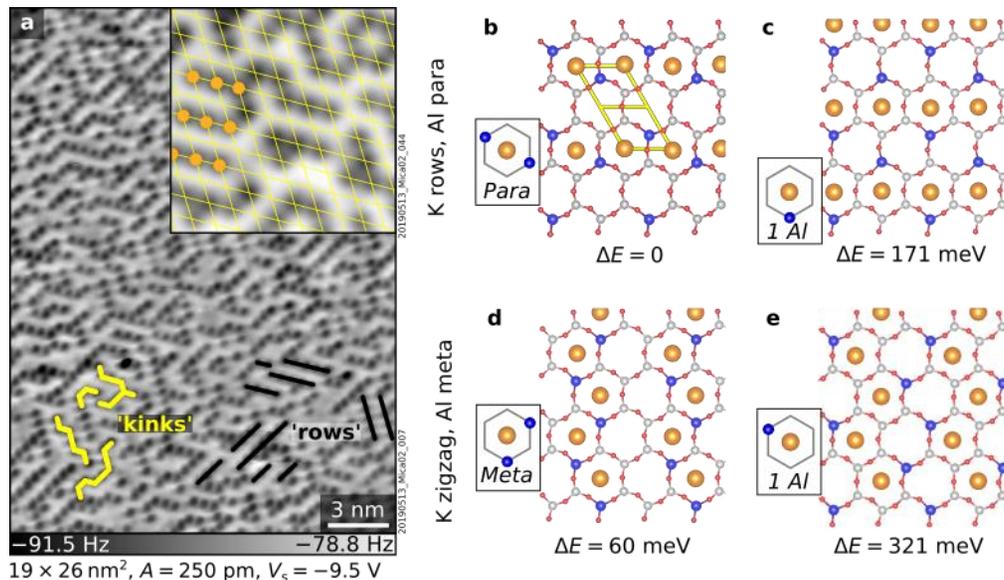

Figure 3. **Distributions of surface K$^+$ and subsurface Al$^{3+}$ ions. a** nc-AFM images of a UHV-cleaved mica surface. Ions arranged in alternating rows and 120° kinks, the most common arrangements, are marked in black and yellow, respectively. The 5 × 5 nm$^2$ inset highlights the hexagonal lattice of mica (yellow) and local K$^+$ order as alternating rows (orange). **b**−**e** DFT-calculated surface structures viewed from above (K$^+$ ions: orange, Al$^{3+}$ ions: blue; Si$^{4+}$ ions: grey; see Supplementary Fig. 1 for the complete set of structures). They differ in the ordering of the surface K$^+$ (rows or zigzag) and subsurface Al$^{3+}$ ions (meta or para), and the number and of the Al$^{3+}$ ions in the K-occupied rings (see insets). Surface energy differences ($\Delta E$) per K$^+$ ion referred to the model in panel **b** are shown at the bottom.

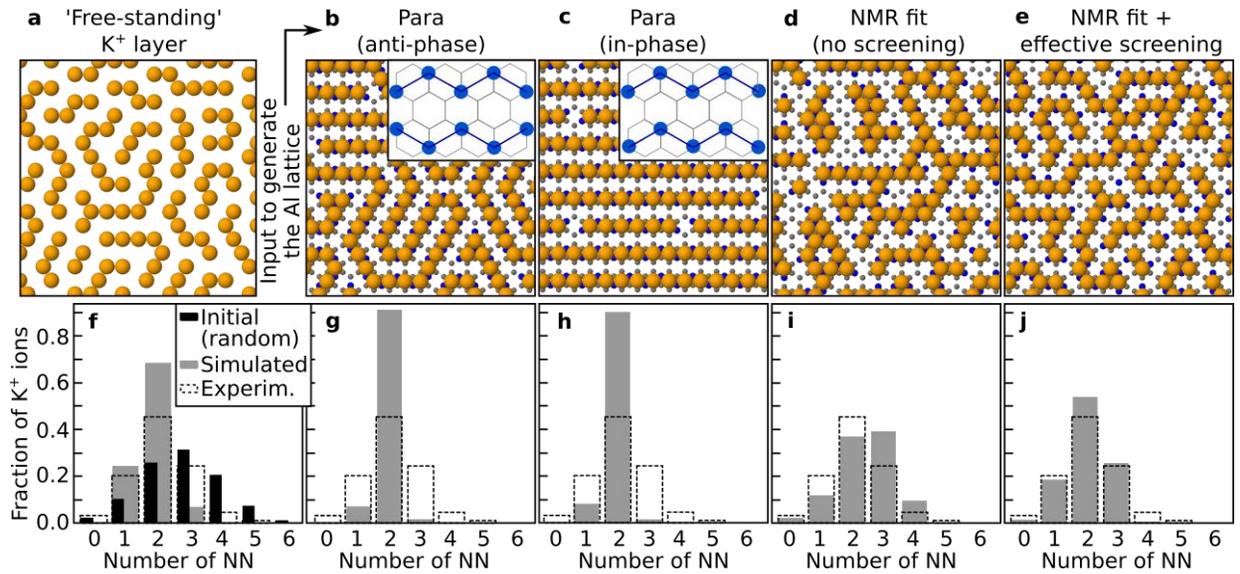

Figure 4. **Monte Carlo simulations of K$^+$ arrangements for different Al$^{3+}$ arrangements.** **a** Distribution of free-standing K$^+$ ions (orange). **b–e** Distributions of K$^+$ ions over different AlSi$_3$ lattices (Al$^{3+}$ and Si$^{4+}$ ions are shown in blue and grey, respectively). **b, c** Long-range-ordered distributions of Al ions obtained assuming perfectly ordered para-only configurations. **d** Al distribution fitting the NMR data satisfying electrostatic constraints on the Al positions (1 or 2 Al/ring, no Al nearest neighbors (NN), meta and para equally favored)[52]. **e** Same Al distribution as in panel **d** but including screening effects through the substrate by considering a lower effective Al charge (details in Supplementary Note 2). **f–j** Histograms showing the fraction of K$^+$ ions found with a given numbers of NN. The distribution extracted from the experimental data (dashed) is overlaid to the simulated ones. A good fit is obtained for the distribution in panels **e, j**.

# Supplementary Information – Resolving the intrinsic short-range ordering of K$^+$ ions on cleaved muscovite mica

G. Franceschi* et al.

*franceschi@iap.tuwien.ac.at

# Supplementary Figures

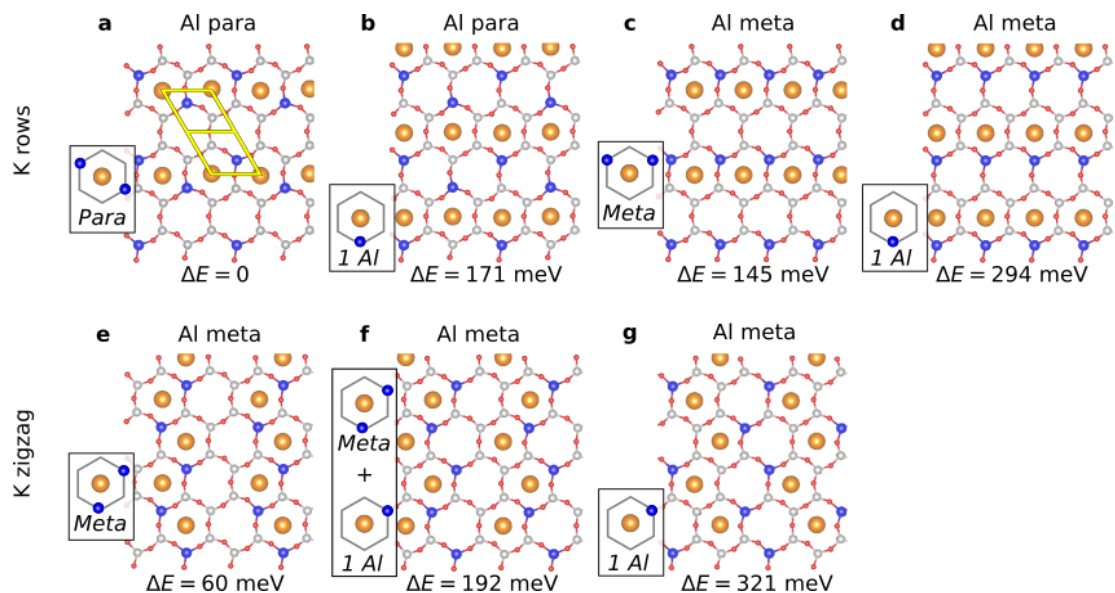

**Supplementary Fig. 1. Distributions of surface K$^+$ and subsurface Al$^{3+}$ ions: full set of structures calculated by DFT. (a−g)** Structures (top view) and energy differences per K$^+$ ion with respect to the lowest-energy structure in panel **a**. Each structure is identified by the K order (rows or zigzag, in the upper and lower row, respectively), the Al order in the substrate (meta or para), and by the position and number of the Al ions in the K-occupied rings (1 or 2, the latter in para or meta configuration – see corresponding insets).

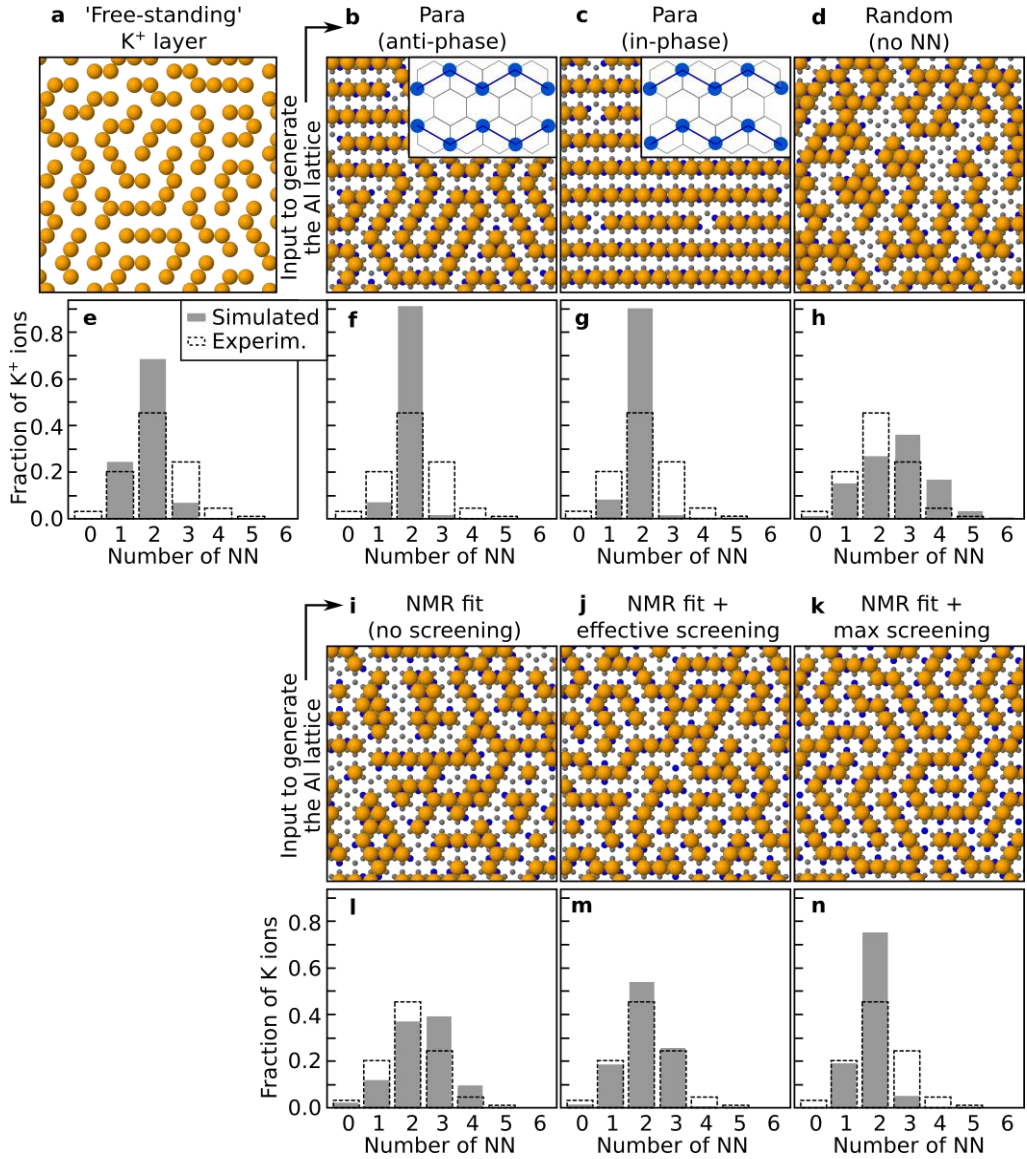

**Supplementary Fig. 2. Full set of Monte Carlo simulations of K distributions for different Al distributions**. **a−d**, **i** Distributions of $K^+$ ions (orange) obtained over $AlSi_3$ lattices (Si: grey, Al: blue) in a purely electrostatic model that considers K and Al as +1 and −1 point charges. **i−k** Distribution of $K^+$ ions over an $AlSi_3$ lattice fitting the NMR data[1] (see main text) and accounting for different screening effects through the substrate (different effective charge of Al, $q_{Al}$): **i** no screening, $|q_{Al}| = 1e$; **j** medium screening, $|q_{Al}| = 0.65e$; **k** strong screening ($|q_{Al}| = 0.24e$) yielding almost the same distribution as the free-standing layer of panel **a**. **e−h, l−n** Histograms representing the fraction of $K^+$ ions found with given numbers of nearest neighbors (NN). The grey bars show the simulation results. Dashed bars correspond to the distributions extracted from the experimental data.

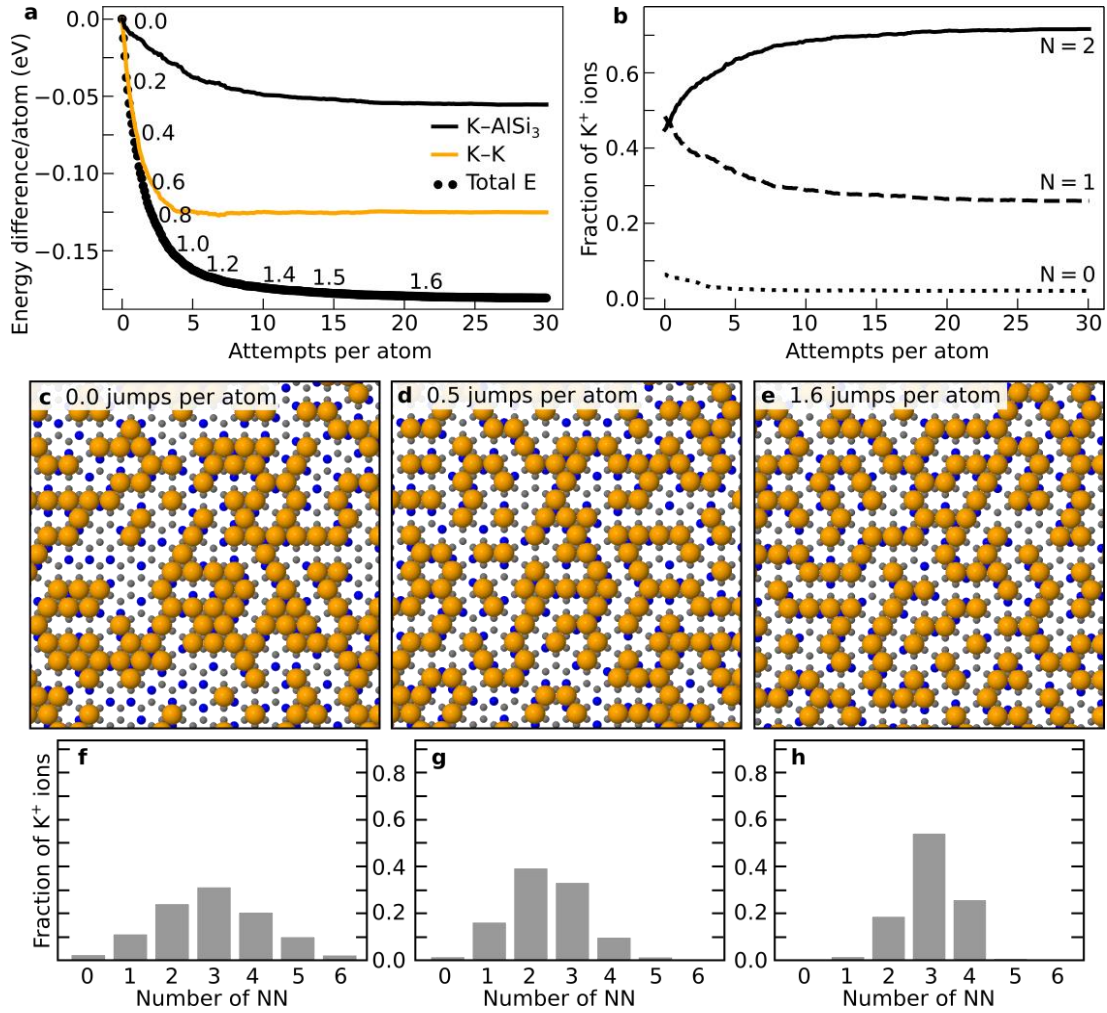

**Supplementary Fig. 3. MC-simulated relaxation of the K ion morphology with the best fit to the experimental data (Fig. 4e in the main text). a** Evolution of the total electrostatic energy and its components. The numbers indicate the realized hops per atoms. The energy values are referred to the initial state. **b** Evolution of the relative numbers of K ions in rings with $N = 0$, $N = 1$, and $N = 2$ Al atoms. **c–e** Snapshots of the morphology at different numbers of attempts per atom. **f–h** Histograms showing the fraction of $K^+$ ions found with a given number of NN for the configurations in panels **c–e**.

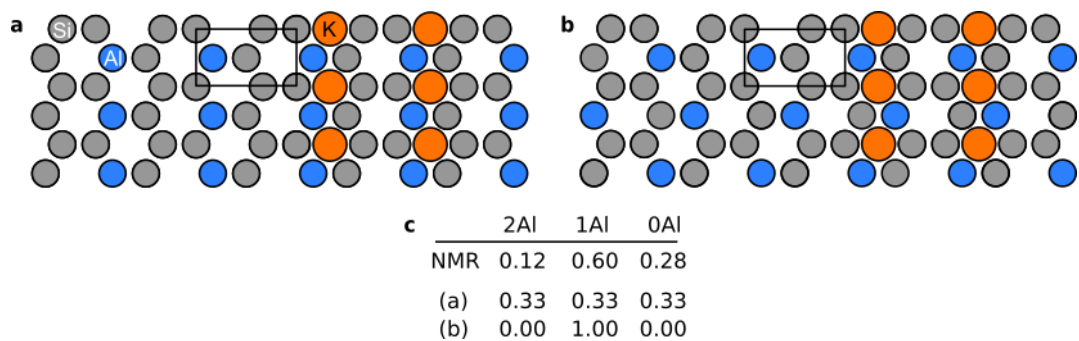

**Supplementary Fig. 4. Structural elements for the Al arrangement in the AlSi₃ layer. a, b** Fully ordered AlSi₃ layers, with Al in only meta, and only para configurations in the rings with 2 Al, respectively. **c** Experimental and calculated probabilities of finding Si atoms with 2, 1, and 0 Al NN. An appropriate mixture of models **a** and **b** can yield the NMR-derived probabilities when accounting for the deviation from perfect 1:3 Al:Si stoichiometry in the tetrahedral sheets. Due to the preference of K for maximizing the number of Al neighbors, the minimum-energy K configuration for these structures would correspond to perfectly ordered K rows.

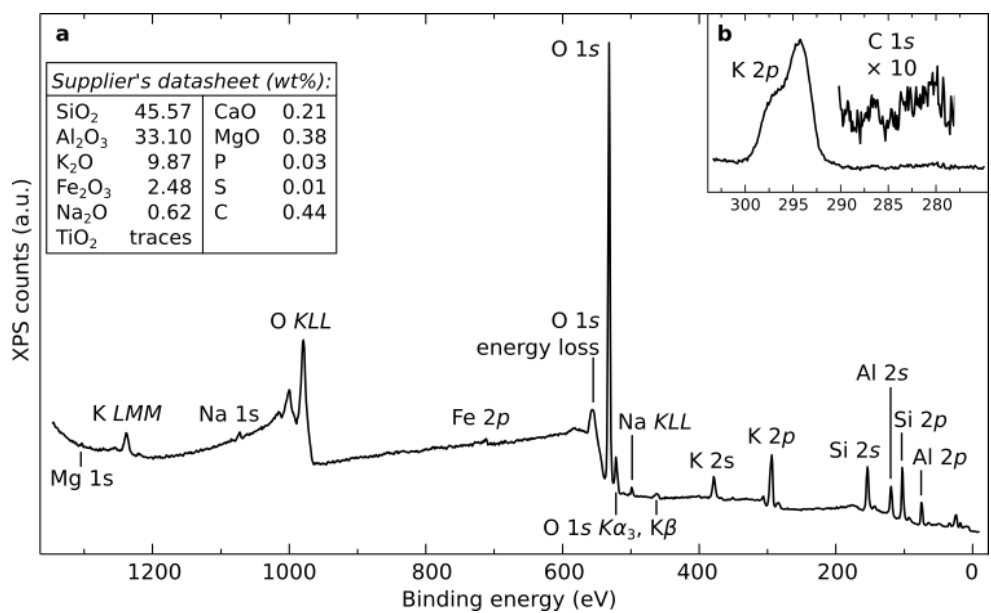

**Supplementary Fig. 5. XPS on the UHV-cleaved surface. a, b** XPS survey and K $2p$+C $1s$ region (after removing the Al K$\alpha_3$ satellite) of an as-cleaved mica surface (Al K$\alpha$, 1486.61 eV, 70° grazing emission, pass energy 60 eV, and 20 eV, respectively). The inset reports the chemical composition from the supplier[2]. The binding energy axes were adjusted to account for charging (see Methods).

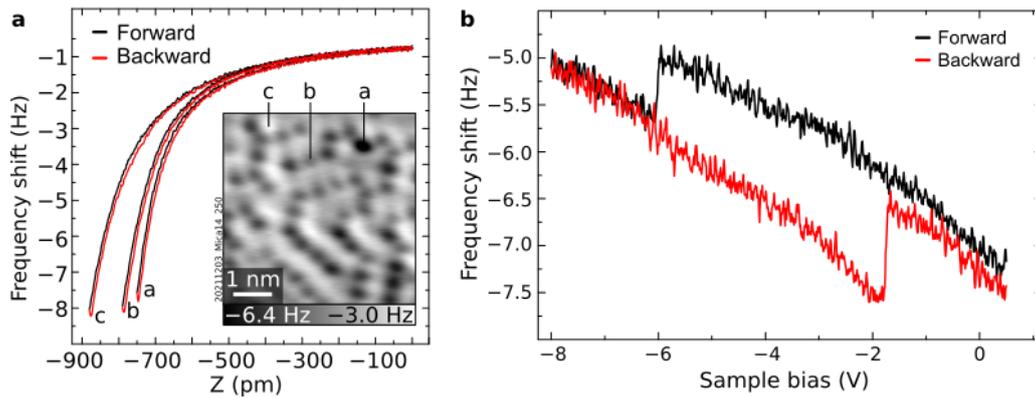

**Supplementary Fig. 6. AFM spectroscopies.** **a** Frequency shift vs. tip-sample distance acquired with a CO tip on three types of features on the cleaved mica surface labeled as "a", "b", and "c". Each curve averages over three runs on the same atom (forward: black, backward: red). $V_s = -10$ V. **b** Typical jumps in the frequency shift during the acquisition of a Kelvin parabola on an as-cleaved mica surface.

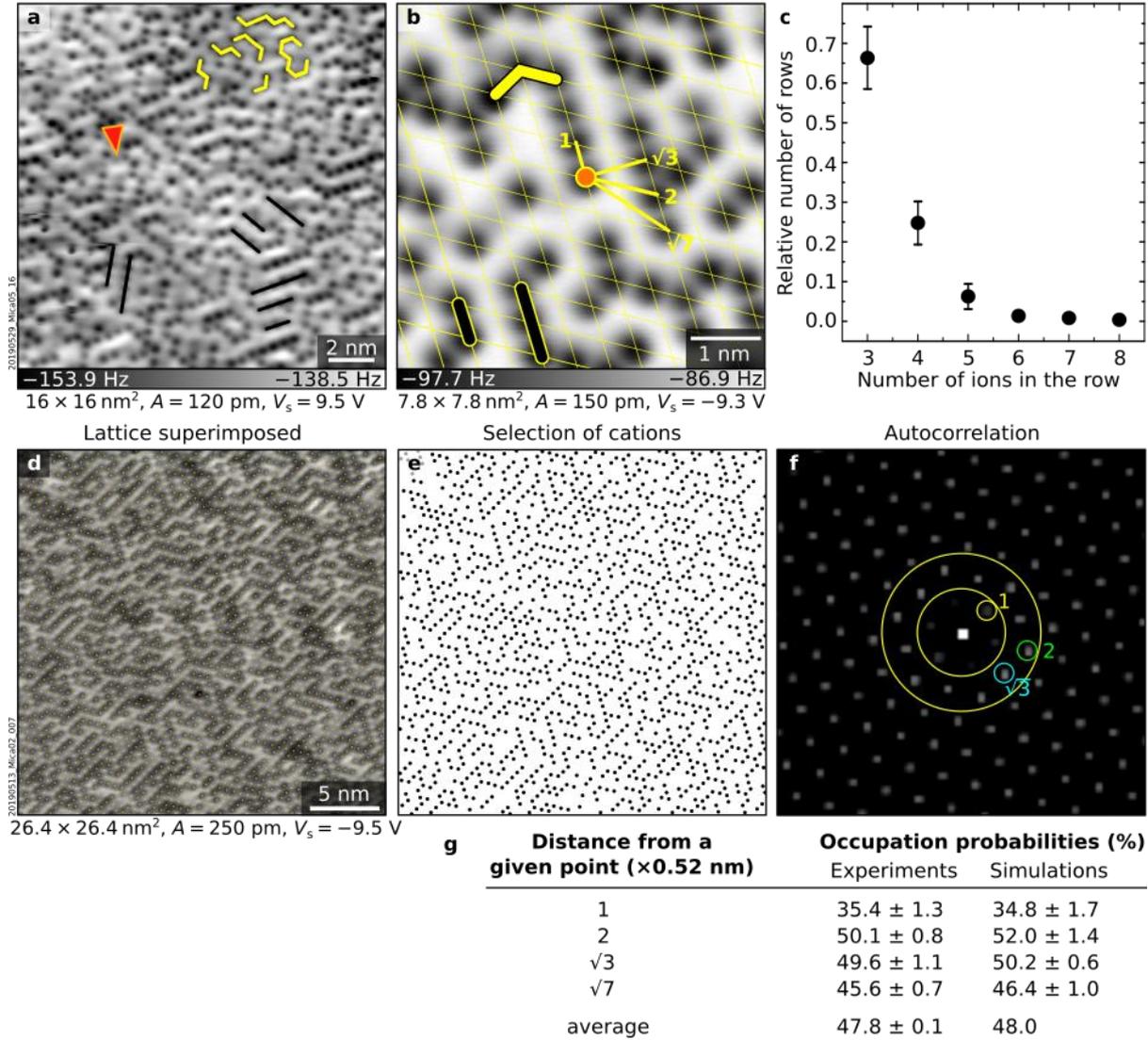

**Supplementary Fig. 7. Statistical analysis of the nc-AFM images. a, b** nc-AFM images highlighting **a** common ion arrangements, i.e., alternating rows and 120° kinks (black and yellow, respectively), and **b** the hexagonal lattice of mica (yellow) with selected ion-ion distances $d$ in units of 0.52 nm. The red arrow in **a** points to a species of fainter contrast, assigned to a $Na^+$ ion. **c** Normalized number of straight ion rows or row sections made of a given number of ions. **d–g** Statistical analysis of the ion positions on the as-cleaved mica surface. **d** nc-AFM image of a UHV-cleaved mica surface with the hexagonal lattice derived from its Fourier transform superimposed (yellow). **e** Lattice of the $K^+$ ions identified with a threshold approach (not accounting for the faint species). **f** Autocorrelation of **e**. Yellow circles aid the eye in identifying regions of different occupation probability (brighter contrasts correspond to higher probabilities). Selected points at $d = 1$, 2, and $\sqrt{3}$ from the reference ion are marked. **g** Occupation probabilities as a function of distance extracted from the experimental data and from the MC simulation of Fig 4e.

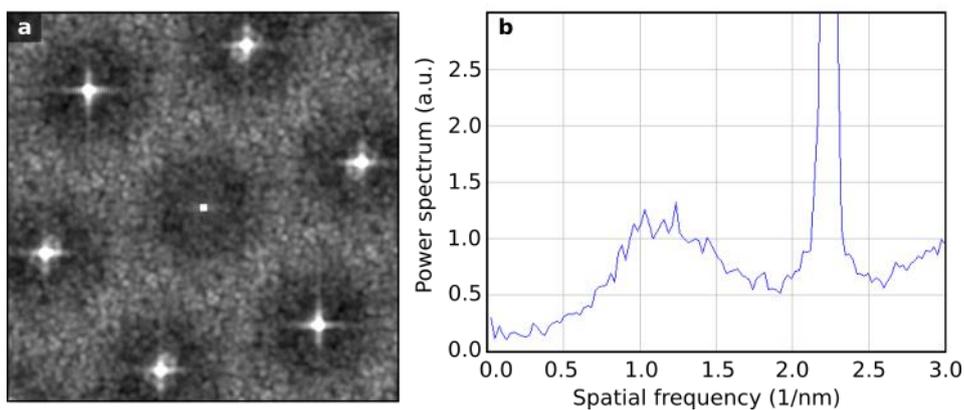

**Supplementary Fig. 8. Fourier transform of the experimental potassium ion distribution.**
**a** Fourier transform of the mask in Supplementary Fig. 6e. The crosses at the strongly overexposed first-order maxima are artifacts caused by the input image being a square. **b** Plot of the diffuse intensity of the power spectrum as a function of the spatial frequency, obtained by averaging over all directions in the power spectrum.

# Supplementary Tables

**Supplementary Table 1:** Calculated relative heights and projected total charges for $K^+$ ions occupying aluminosilicate rings characterized by 0, 1, 2, or 3 Al atoms.

| # of Al atoms in the occupied ring | Relative height (Å) | Projected total charge ($e$) |
|---|---|---|
| 0 | 1.84 | 8.01 |
| 1 | 1.89 | 8.03 |
| 2 | 1.84 | 8.03 |
| 3 | 1.97 | 8.05 |

# Supplementary Notes

## Supplementary Note 1: Additional details on the Density Functional Theory (DFT) calculations

**Full set of tested DFT structures**

Supplementary Figure 1 shows the complete set of DFT-tested structures, highlighting the order of the K and Al ions, the arrangement of the Al ions in the K-occupied rings, and the energy per $K^+$ ion referred to the lowest-energy structure of Supplementary Fig. 1a.

**Relative height and projected total charge for $K^+$ ions in 0, 1, 2, 3 Al rings**

Supplementary Table 1 summarizes the relative heights and projected total charges of $K^+$ ions positioned in different aluminosilicate rings. The values were calculated after fully relaxing the ionic positions in a single trilayer slab. They were almost identical for a double trilayer slab (differences <0.01 Å and <0.01$e$ for the ionic positions and the projected total charges, respectively).

The relative heights were calculated as the difference between the z coordinate of the surface $K^+$ ion and the average of the z coordinates in the $AlSi_3$ subsurface plane. All values are very similar, except for the case of 3 Al atoms in the occupied ring. This arrangement is considered a defect, since it does not comply with the principle of maximum charge dispersion[1]. Thereby, the predicted height difference of ≈0.1 Å compared to the non-defective rings is expected to yield a visible difference in the nc-AFM images. It is possible that the very dark species sparsely seen in the images correspond to $K^+$ ions sitting on these sites. Another possibility is that these darker species correspond to substitutional $Ca^{2+}$ species (see Supplementary Note 5 below).

The projected total charges were calculated as the sum of the projected partial charges of the $s$, $p$, and $d$ orbitals within the PAW sphere, setting LORBIT=11 (VASP version 6.3.0). The projected total charge for a K ion in the bulk is equal to 8.00$e$, similar to the values for surface K ions in Supplementary Table 1.

## Supplementary Note 2: Additional details on the Monte Carlo simulations

As mentioned in the Methods Section, the MC simulations shown in this work stem from a simplified model that only considers atom-atom electrostatic interactions with a 1/r potential. Other factors that might play a role, e.g., covalent bonding contributions and variations of lattice relaxations at the K sites, have not been included for the following reasons: The influence of covalent bonding contributions should be limited, as the bonding of K is expected to be mostly ionic and the variation of the covalent bonding strength among different K sites should be small. Also lattice relaxations, both out-of-plane and in-plane, should not be critical; simulations assuming the bulk interlayer distance of 2.28 Å between the K layer and the tetrahedra sheet (not shown) are very similar to the simulations in Fig. 4 and Supplementary Fig. 2. There, the distance obtained from the DFT-relaxed cleaved surface (1.85 Å) was assumed. In the DFT-simulated structures, the in-plane displacements of the K ions compared to the ideal lattice sites is less than 4% of the K–K NN distance. This suggests that lattice relaxations should not play a significant role for the arrangement of the $K^+$ ions. The largest error of the simulation energies probably comes from approximating the screening of the electrostatic interactions by the substrate, approximated by a reduced effective charge on the Al. This neglects the distance-dependence of the screening, and the screening of the K–K interactions. Developing a high-accuracy screening model would be difficult, as some of the required parameters – e.g., the layer-resolved anisotropic dielectric constant at the timescale of single diffusion events – are not known. In any case, the simple model used here should be sufficient, as the simulations are mainly determined by the strong interactions of nearby ions and insensitive to quantitative values: They only depend on whether a diffusion step is uphill or downhill in energy.

Supplementary Figure 2 shows the complete set of MC-simulated structures, i.e., the results presented in Fig. 4 plus additional simulations. The simulations were obtained with different distributions of the Al atoms in the subsurface $AlSi_3$ sheet. Supplementary Figures 2b, c show long-range ordering according to the corresponding insets. Supplementary Figure 2d is based on an Al distribution with no nearest neighbors (NN), according to the known Löwenstein's rule for aluminosilicates[3], a maximum of 2 Al ions/ring, but otherwise a random arrangement. Supplementary Figure 2i is based on an Al arrangement compatible with the NMR data in ref.[1] and satisfying electrostatic constraints on the Al positions, i.e., no NN, maximum 1 or 2 Al/ring, and random placement apart from these constraints. The meta:para ratio obtained with these constrains corresponds to 2:1[1]. Supplementary Figures 2j, k are based on the same Al distribution of Supplementary Fig. 2i but include an approximation for screening effects through the substrate to depict a more realistic scenario. The effect of screening is simulated by reducing the formal charge of the Al ions and spreading the missing charge over the Si ions in the $AlSi_3$ layer to retain a charge-neutral system. Considering a weaker charge on the Al sites is reasonable: When considering only NN interactions, DFT indicates that the K−K and the K−Al interactions are comparable in energy (≈0.15 eV and 0.17 eV, respectively), despite the larger distance of K−K neighbors (0.52 nm) compared with K−Al (0.36 nm). Hence, the K−Al interaction should be described well by a weaker charge of Al. The effective charge of $-0.65e$ chosen for the distributions in Supplementary Figs. 2j (reprinted in Fig. 4e of the main text) yields the best fit with the experimentally observed K arrangements.

Supplementary Figure 3 demonstrates the evolution of the K ion morphology during the simulation using the subsurface Al configuration compatible with NMR and the Al effective charge best fitting the experiment, i.e., the setup of Fig. 4e and Supplementary Fig. 2j. Supplementary Figure 3a shows the evolution of the total electrostatic energy of the configuration and its contributions. The most prominent contribution is the relaxation due to K–K repulsion (solid orange line), much stronger than the K–Al contribution (solid black line). The numbers in the graph indicate the average number of jumps per atom to reach a given energy. Supplementary Figures 3c−e depict the evolution of the morphology: Panel (c) shows the initial random configuration, while panels (d) and (e) correspond to the configurations after 0.5 and 1.8 realized jumps per atom, respectively. Supplementary Figures 3f−h show the corresponding histograms of NN K ions. Supplementary Figure 3b presents the evolution of the correlation between the K ions and the subsurface Al ions. It shows the relative numbers of K ions in rings containing $N = 0$, $N = 1$, and $N = 2$ Al atoms. A tendency to occupy rings with 2 Al ions is evident.

**Comment on additional Al arrangements in the AlSi$_3$ layer**

In this work, it was shown that the measured distribution of K$^+$ ions can be fitted well by considering the (screened) interaction with a subsurface AlSi$_3$ layer that satisfies electrostatic constraints on the Al positions and fits previous NMR data[1]. This model, referred to as NMR fit, and based on a statistical placement of the Al atoms with the posed constraints, is not unique, however. Here it is argued that other solutions with more pronounced ordering fit the NMR data; these may influence the surface K$^+$ ordering.

Consider the two fully ordered AlSi$_3$ layers in Supplementary Fig. 4, with Al ions in (a) meta and (b) para positions. An appropriate mixture of structures would fit the NMR data: The first row in panel (c) shows the experimental probabilities for Si ions in different environments, i.e., with 2 Al, 1 Al, and 0 Al NN. The underlying rows yield the probabilities derived from the fully ordered models in (a) and (b). Considering that the difference between the 2 Al and 0 Al values in NMR are only due to an Al deficiency (Al:Si concentrations in the tetrahedral sheet

< 1:3), a good match with experiments may be achieved by appropriately mixing the two structures. The structures in Supplementary Figs. 4a, b would favor alternating $K^+$ rows. In other words, a mixture of these fully ordered models can be compatible with the NMR data and, at the same time, favor $K^+$ rows. Clearly, the fully ordered models considered here are extreme scenarios not occurring in reality. Nonetheless, this hypothetical structure indicates that structures with local ordering are compatible with the NMR data and, at the same time, favor short-range ordering of the $K^+$ ions exceeding that observed in the experiments.

## Supplementary Note 3: XPS survey spectrum and C 1*s* region of UHV-cleaved mica

Supplementary Figure 5a shows an XPS survey acquired in grazing emission on a UHV-cleaved mica surface. The spectrum is consistent with others found in the literature[4,5], showing trace amounts of Mg, Fe, and Na besides the elements characteristic of mica (O, K, Si, and Al). The inset in panel (b) shows the K 2*p* peak and the adjacent C 1*s* region after removing the Al $K_{\alpha 3}$ satellite. The sample is virtually free from carbon.

## Supplementary Note 4: AFM-based spectroscopies

Supplementary Figure 6 shows two AFM-based spectroscopies performed on UHV-cleaved mica surfaces: frequency shift vs. tip-sample distance and frequency shift vs. sample bias voltage (Supplementary Figs. 6a, b, respectively).

Supplementary Figure 6a plots curves acquired on representative features identified based on their contrast: Features labeled as "a" are the round species appearing darker than average $K^+$ ions. Black dots assigned to "regular" $K^+$ ions are labeled as "b", and white regions in the background as "c". Before each acquisition, the tip was positioned on the chosen feature and retracted by 500 pm from the acquisition height of the image in the inset. In Supplementary Fig. 6a, the same value of frequency shift occurs at different tip-sample distances for the three different features: The biggest tip displacement towards the sample is required in region "c" between the $K^+$ ions, followed by features "b" and then "a". This indicates that darker features protrude more than brighter ones or that they attract the tip more strongly due to a higher local charge or different chemical contrast. Note that the ions were almost exclusively measured in the attractive regime regardless of the tip termination, even at extremely close distances. Hence, a minimum in the curves of Supplementary Fig. 6a could not be observed. The exclusively attractive interaction between isolated, undercoordinated cationic adatoms on the surface and the tip is typical in nc-AFM[6,7]. Note also that the curves are quantitatively reproducible only within regions of a few nanometers that show the same background contrast. Quantitative values vary over distances of a few nanometers because the charges introduce a significant contribution of long-range forces to the overall tip-sample interaction.

$\Delta f(V)$ curves (Kelvin parabolas), used to calculate the local contact potential difference (LCPD), were acquired on features "a", "b", "c" of Supplementary Fig. 6a on different UHV-cleaved mica surfaces (before acquiring each curve, the tip was positioned on the chosen feature and retracted by 500 pm from the height at which the image in the inset was acquired). The general trend is that Kelvin parabolas acquired on more attractive features have their maxima shifted in sample bias by roughly −2 V compared to Kelvin parabolas taken on the background, indicating that the attractive features are positively charged, as expected for $K^+$. Unfortunately, quantitative conclusions cannot be drawn: Many curves appear like in Supplementary Fig. 6b, with one or two sudden jumps in the frequency shift. After such a jump, the Kelvin parabola acquired on the same position is shifted in bias compared to the previous one. The behavior is interpreted as charging/discharging events. Ion manipulation can be excluded since images acquired before and after the jump events show the same arrangement of the K ions. The smaller the tip-sample distance, the more probable the events. As for the $\Delta f(z)$ curves discussed above, quantitative reproducibility is hindered by the presence of regions with different long-range interactions with the tip.

## Supplementary Note 5: Analysis of the nc-AFM results

Supplementary Figures 7a, b show nc-AFM images of UHV-cleaved mica while highlighting common ion arrangements on UHV-cleaved mica, i.e., alternating rows (black) and 120° kinks (yellow) that often interrupt or connect rows. Supplementary Figure 7b shows a detailed view of the mica structure, the hexagonal lattice (yellow), and some relevant distances $d$ in units of 0.52 nm. The separation between the two ions next to a 120° kink is $d=\sqrt{3}$.

**Counting $K^+$ ions**

Supplementary Figures 7d−f illustrate the typical analysis performed via the software ImageJ[8] to count and evaluate the distribution of the $K^+$ ions on UHV-cleaved mica surfaces. The analysis was performed on ≈10 images from different samples with similar image quality to Supplementary Fig. 7a. Image sizes ranged from ≈15 × 15 $nm^2$ to ≈40 × 40 $nm^2$, corresponding to an absolute number of counted features per image between 600 and 3600. Each nc-AFM image was first undistorted based on its Fourier Transform[9]. The reference lattice obtained

from a back transform of the first-order Fourier spots was superimposed on the undistorted nc-AFM image, as shown in Supplementary Fig. 7d. In Supplementary Fig. 7d, all ions (black dots) sit on the reference lattice (yellow). The ions were selected and counted with a threshold approach, producing the mask shown in Supplementary Fig. 7e.

This analysis yielded a coverage of ≈46% when counting uniformly dark species, and 47.8±0.1% when selecting also fainter species such as the one highlighted by a red arrow in Supplementary Fig. 7a and Figs. 1c, d (errors in the statistical evaluations of AFM images represent 95% confidence intervals calculated with a two-tailed Student's t-distribution from the standard error of the mean). This value is smaller than the 50% expected for perfectly stoichiometric samples. The most likely reasons for this discrepancy are: (i) An Al:Si ratio higher than 1:3 in the tetrahedral sheets, which would require a lower K concentration in the K layer to achieve charge neutrality. Such a deviation from the ideal stoichiometry was already reported for the samples used observed by Herrero et al.[1] (3.16 and 0.84 for Si and Al in the tetrahedral sheet, and 0.84 in the K interlayer, which contained also trace amounts of $Ca^{2+}$) and is consistent with the datasheet of the samples used here (see Supplementary Fig. 5). (ii) Presence of +2 impurities replacing the surface $K^+$. In this case, fewer ions would be needed at the surface because the same negative charge can be compensated by half the number of +2 species compared to +1 species. These species could be $Ca^{2+}$ ions, which are present according to the supplier's datasheet, and are supposed to substitute $K^+$. Because of their higher charge, these species should appear darker than $K^+$ ions (the average features) in nc-AFM. In fact, some species darker than average are recognized in the images (arrows highlight examples in Figs. 1c,d of the main text). Nonetheless, their concentration is too small to account for the difference between 47.8 and 50%. As discussed in Supplementary Note 1, these darker species could also correspond to $K^+$ ions sitting on (rare) aluminosilicate rings with three Al sites. Among the sample's other impurities, $Fe^{2+}$ and $Mg^{2+}$ do not substitute $K^+$, but $Al^{3+}$ in the octahedral sheets[10]. $Na^+$ species replace $K^+$ and should appear fainter in nc-AFM because of their smaller ionic radius (see below for more comments on $Na^+$ ions). Other possibilities for the less-than-50% surface ion coverage are (iii) Missing $OH^-$ groups from the subsurface; in this case, fewer $K^+$ ions would be needed for a charge-neutral system. (iv) Errors in the analysis of the nc-AFM images. In the images, ≈3−5% of the dark dots appear with fainter contrast than the rest (see the red arrow in Supplementary Fig. 7a for one example). It is not unreasonable that some of these faint species might be missed during the statistical evaluation. These species are likely to be $Na^+$ ions: They are expected to substitute $K^+$ and appear with fainter contrast due to their smaller ionic radius compared to $K^+$. Moreover, the nominal Na/(Na+K) ratio according to the datasheet in Supplementary Fig. 5 is 8.7%, in the same order of the number of faint species counted in the nc-AFM images.

**Average length of rows and row sections**

Supplementary Figure 7c shows an analysis of the length of the straight ion rows (including straight sections of rows with kinks; excluding rows made of 2 ions; normalized to the number of rows in each image; error bars represent 95% confidence intervals calculated with a two-tailed Student's t-distribution from the standard error of the mean). Most of the straight rows or straight row sections are made of 3 $K^+$ ions (average excluding the rows made of 2 ions: 3.5±0.4 ions).

**Autocorrelation analysis**

An autocorrelation image (Supplementary Fig. 7f) was obtained from the mask of Supplementary Fig. 7e, showing the cations with normalized contrast. The autocorrelation image shows the probabilities of finding ions at given relative distances (see averaged probabilities in Supplementary Fig. 7g). In Supplementary Fig. 7f, regions of different occupation probabilities are marked by yellow circles to aid the eye. The six nearest neighbors ($d = 1$, ring closest to the center) have an occupation probability of 35.4±1.3%, lower than the average of 47.8±0.1%. Instead, the occupation probability of the ions at $d = 2$ and $d = \sqrt{3}$ (second ring) is slightly higher than average (49.9±1.4%). For ions farther away, it becomes close to average. The measured depletion at the nearest-neighbor positions is due to repulsion between the $K^+$ ions, which leads to alternating straight or kinked rows. These preferred geometries also lead to the probabilities of finding ions at $d=2$ and $d=\sqrt{3}$ greater than average. Assuming a perfect arrangement of infinite alternating rows (50% $K^+$ concentration), each ion within a row will have two out of the six nearest-neighbor sites occupied. The same is true for alternating zigzag (kinked) rows as shown in Fig. 3f of the main text. Thus, the probability of finding a $K^+$ nearest neighbor would be 1/3. The probability of finding a $K^+$ neighbor at $d = \sqrt{3}$ would be 1/3 and 2/3 for the ideal straight and zigzag configuration, and at $d = 2$ these probabilities would be 1 and 1/3, respectively. Since the experiments show both straight and kinked arrangements, the probabilities for $K^+$ neighbors at both $d = \sqrt{3}$ and 2 are between these two ideal cases. Since only short-range ordering is present, these probabilities are not far from the $K^+$ concentration. Notice that the occupation probabilities extracted from the experimental data fit reasonably well with those obtained from the MC-simulated distribution of Fig. 4e of the main text (see Supplementary Fig. 7g).

**Comments on the Fourier Transform of the nc-AFM images**

Notice that the background of the Fourier transform in Fig. 1e in the main text is not uniform, as it would be for a purely random distribution of $K^+$ ions. It displays a sort of star with maximum intensity within a rim at half the unit cell distance and arms stretching between the first-order periodicities. To unveil the origin of this background, the Fourier transform of the mask shown in Supplementary Fig. 7e was calculated (see Supplementary Fig. 8a; Supplementary Fig. 8b shows the corresponding power spectrum as a function of the spatial frequency, averaged over all azimuthal directions). This Fourier transform reproduces the features of the original nc-AFM image (apart from the decay towards high spatial frequencies, which is due to the finite lateral resolution of the nc-AFM image). Thus, the uneven background must originate from the positions of the ions and not from modulations in the AFM background or imaging artifacts.

The higher intensity within the rim between the zero- and first-order spots, also seen in the power spectrum of Supplementary Fig. 8b as a shallow and wide peak at spatial frequencies around 1 nm$^{-1}$, is because of the alternating rows formed at the surface (separation of 2 unit cells in real space). The peak is neither high nor sharp because of the lack of long-range order.

The depletion of the intensity close to the origin (and around the higher-order Fourier spots, which are essentially replicas of the origin) shows that the system avoids long-range coverage fluctuations, as expected for distributions dictated by electrostatic interactions. This interpretation is consistent with the fact that a similar Fourier transform is also obtained for the simulated $K^+$ lattice assuming electrostatic repulsion in the MC results in Fig. 4a (not shown). The ion distribution of Fig. 4e is different from that of Supplementary Fig. 7e, but it is still determined by electrostatic interactions. Hence it also minimizes long-range fluctuations in the local charge distribution.

In principle, one would expect to observe the variations of the diffuse background seen in the Fourier transform also in the LEED images. However, the additional diffuse phonon background around the diffraction spots[11] produces an increased intensity around the spots, which may hide the diffuse background caused by the $K^+$ distribution. This is probably the reason why early LEED works did not infer any hint of short-range order from their data[12].

## Supplementary Note 6: Comment on LEED

Acquisition of LEED was attempted on the UHV-cleaved surface without success. The samples charge immediately when irradiated with the LEED electron beam. Patterns consistent with the ones obtained by Müller et al.[12] could be obtained on samples that were not pristine, e.g., left in the UHV chamber for >24 h or exposed to small amounts of water at low temperature and warmed up to room temperature again. Apart from the trace elements already present after cleavage, the latter samples appeared clean within the detection limit of XPS[12].

# Supplementary References